\documentclass{iau} 
\usepackage{graphicx}

\title[Multiple populations in GCs: kinematics and dynamics] 
{Multiple populations in globular clusters: \\ constraints from kinematics and dynamics}

\author[Vincent H\'enault-Brunet]   
{Vincent H\'enault-Brunet}

\affiliation{Department of Physics, University of Surrey,\\ GU2 7XH, Guildford, UK 
 \\ email: {\tt v.henault-brunet@surrey.ac.uk} 
}

\pubyear{2015}
\volume{316}  
\jname{Formation, Evolution, and Survival of Massive Star Clusters}
\editors{C. Charbonnel \& A. Nota, eds.}
\begin{document}

\maketitle

\begin{abstract}
We discuss constraints on the formation of multiple populations in globular clusters (GCs) imposed by their present-day kinematics (velocity dispersion and anisotropy) and spatial distribution. We argue that the observational evidence collected so far in the outer parts of clusters is generally consistent with an enriched population forming more centrally concentrated compared to the primordial population, in agreement with all the scenarios proposed to date (in some cases by design), but not sufficient to favour a particular scenario. We highlight that the differential rotation of subpopulations is a signature that may provide crucial new constraints and allow us to distinguish between various scenarios. Finally, we discuss the spatial distribution of subpopulations in the central regions of GCs and speculate that mass segregation between subpopulations may be due to a difference in their binary fraction.

\end{abstract}

\firstsection 
\section{Introduction}

It is convenient to split the problem of multiple populations in globular clusters (GCs) into two challenges or questions. First, what is producing the chemical abundance anomalies giving rise to observable features such as the Na-O anti-correlation and multiple sequences in the colour-magnitude diagrams of GCs (e.g. Gratton et al. 2012, Piotto et al. 2012)? But another important challenge is to understand how this polluted material makes its way into a large fraction of the cluster stars (the observed enriched fraction is $\sim68\%$ today; see e.g. Bastian \& Lardo 2015). Whether it is through multiple generations of star formation (e.g. D'Ercole et al. 2008), via accretion onto low-mass pre-main sequence stars born as part of the same burst of star formation (Bastian et al. 2013), or via a yet unknown alternative process (as suggested by Bastian et al. 2015), the mechanism by which enriched material is incorporated into stars will dictate the initial configuration of the enriched (i.e. polluted) stars compared to the primordial (i.e. normal) stars. This will set the initial conditions for the long-term dynamical evolution of the cluster and its multiple populations, which is the focus of this contribution.

Because GCs are collisional systems, their long-term dynamical evolution leads to two phenomena that can affect the present-day kinematics and spatial distribution of subpopulations: phase-space mixing and mass segregation. Due to two-body encounters, populations of stars with different initial phase-space distributions will exchange energy and angular momentum and gradually mix (Decressin et al. 2008, Vesperini et al. 2013, H\'enault-Brunet et al. 2015). The trend towards kinetic energy equipartition also leads massive stars and remnants to slow down and sink towards the centre of the cluster, effectively pushing the lower-mass stars outwards. Because the two-body relaxation timescale (on which these processes become important) is inversely proportional to the local density (e.g. Spitzer 1987), the importance of these processes will vary significantly between different regions of the cluster. In the outer regions, where the relaxation timescale is longer, there will be less mixing and we may expect to still see imprints of the initial conditions and formation process. In the inner regions, the shorter relaxation timescale leads to more mixing, potentially erasing any initial differences between subpopulations, and mass segregation can also proceed more efficiently and introduce differences in the spatial distribution and kinematics of different mass species. In the following sections, we review constraints on the formation of multiple populations imposed by their present-day kinematics and spatial distribution.

\section{Spatial distribution \& kinematics of subpopulations in the outer regions of GCs}

N-body simulations of clusters with two subpopulations starting spatially segregated and with comparable numbers of stars have shown that complete spatial mixing is expected when about $60-70\%$ of the initial mass of the cluster has been lost due to two-body relaxation (Vesperini et al. 2013). This also applies to the distinct kinematics of different subpopulations, which will be erased on a similar timescale (e.g. Decressin et al. 2008, H\'enault-Brunet et al. 2015). Given their current mass and galactocentric radius, a large fraction of Galactic GC are expected to still be expanding within their tidal radius and to have lost less than 50\% of their initial mass by relaxation-driven evaporation (Gieles et al. 2011), so some memory of the initial configuration of subpopulations should be preserved in the outer parts of many GCs. And indeed, when looking at the spatial distribution of subpopulations in the outer regions of clusters (from around the half-mass radius - $r_{\rm h}$ - and beyond), a more centrally concentrated enriched population has been found in many clusters (e.g. Sollima et al. 2007, Bellini et al. 2009, Lardo et al. 2011, Milone et al. 2012), suggesting that enriched stars formed more centrally concentrated. That said, evidence for a fully mixed cluster has also been reported (NGC 6362; Dalessandro et al. 2014) and full mixing may be more common than previously thought (Vanderbeke et al. 2015; although see their caveats about using Horizontal Branch stars as tracers).

\begin{figure}[!h]
\begin{center}
 \includegraphics[width=5.2in]{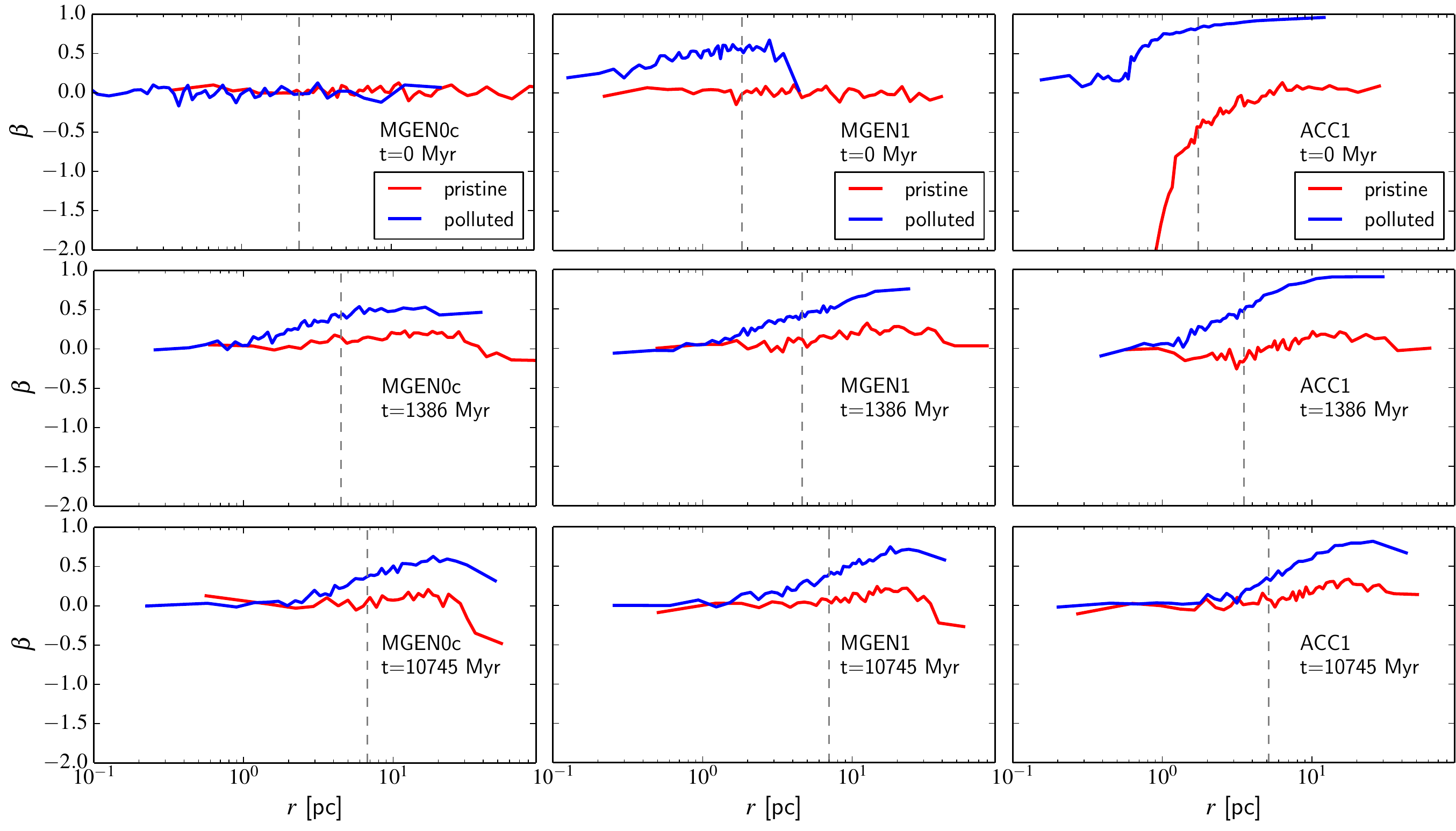} 
 \caption{Velocity anisotropy ($\beta$) as a function of radius (in 3D) for the polluted (in blue) and pristine (i.e. primordial composition - in red) stars as a function of time, for three different initial configurations: a spherical and isotropic subsystem of polluted stars embedded in a more extended spherical cluster (model MGEN0c - left), a flattened rotating subsystem of polluted stars embedded in a non-rotating spherical cluster (model MGEN1 - centre), a rotating cluster where the polluted stars are flagged as the ones with an orbit crossing the cluster core at $t=0$ (model ACC1 - right). The half-mass radius is indicated by grey dashed lines. For more details on these simulations see H\'enault-Brunet et al. (2015), from which this figure is adapted.}
   \label{fig1}
\end{center}
\end{figure}

Kinematic differences have been found in some cases in the outer regions of clusters, with the (Na) enriched population having a lower velocity dispersion (Bellazzini et al. 2012, Ku{\v c}inskas et al. 2014). The more centrally concentrated population is generally expected to be dynamically cooler (see e.g. Bekki 2011; Mastrobuono-Battisti \& Perets 2013; H\'enault-Brunet et al. 2015), regardless of its initial configuration (flattened disc-like or spherical subsystem), so again this is consistent with an enriched population that started more centrally concentrated. Using Hubble Space Telescope (HST) proper motions of stars near $2\,r_{\rm h}$ in 47 Tuc, Richer et al. (2013) found that the presumably He-enriched stars (bluer main-sequence stars) have a more radially anisotropic velocity distribution. Bellini et al. (2015) observed a similar behaviour for stars between $\sim1.5$ and $2 \ r_{\rm h}$ from HST proper motions in NGC 2808, with stars coinciding with the presumably He-enchanced populations in the middle and blue main sequences showing a smaller tangential velocity dispersion than the populations corresponding to primordial stars, while no significant differences were found in the radial-velocity dispersion. We showed in H\'enault-Brunet et al. (2015) that a more radially anisotropic velocity distribution for the polluted stars is also predicted if this population started more centrally concentrated and then diffused outward, and it is by no means a unique signature of any of the scenarios proposed. For example, whether gas collected into the core of the cluster to form a new generation of polluted stars (possibly adopting a flattened initial distribution if the cluster had some net rotation; e.g. D'Ercole et al. 2008, Bekki 2010), or if the polluted stars are the ones initially crossing the core of the cluster and preferentially on radial orbits (e.g. as in the early disc accretion scenario of Bastian et al. 2013), any remaining present-day velocity anisotropy signature is expected to be very similar (see Fig.~\ref{fig1}).

\section{Differential rotation as a potential way to distinguish between formation scenarios}

\begin{figure}[h]
\begin{center}
 \includegraphics[width=3.9in]{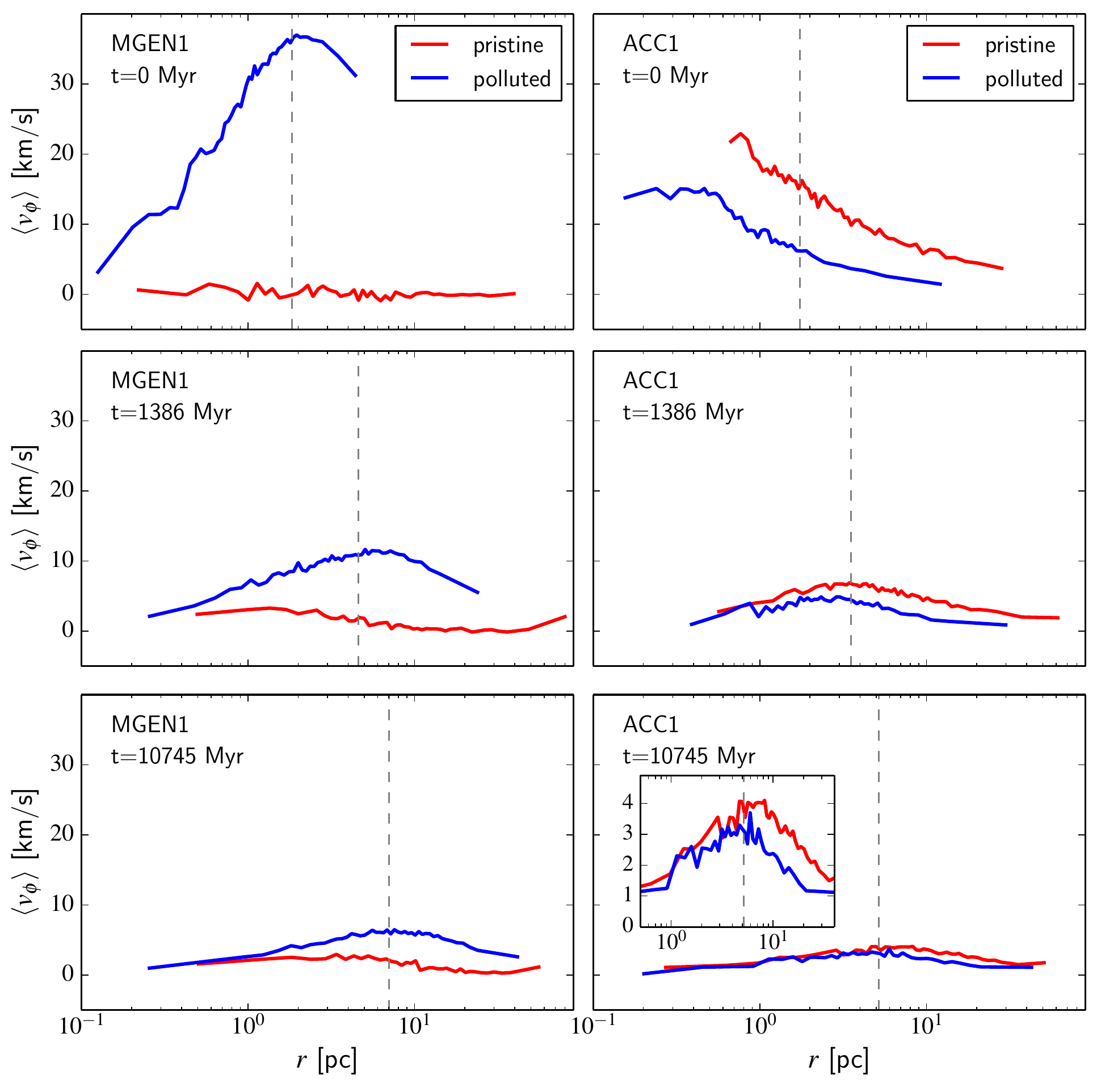} 
 \caption{Mean azimuthal velocity as a function of radius - i.e. rotation curve - (in 3D) for the polluted (blue) and primordial (red) stars as a function of time, for two of the models also considered in Fig.~\ref{fig1}: a flattened rotating subsystem of polluted stars embedded in a non-rotating spherical cluster (model MGEN1 - left), a rotating cluster where the polluted stars are flagged as the ones with an orbit crossing the cluster core at $t=0$ (model ACC1 - right). Adapted from H\'enault-Brunet et al. (2015).}
   \label{fig2}
\end{center}
\end{figure}

All the observational evidence collected so far in the outer parts of GCs points to an enriched population forming more centrally concentrated, but this does not help to discard or favour any particular formation mechanism since all current scenarios either predict such a trend or were designed to match these observational constraints. To make progress, we need to identify unique imprints that would allow to distinguish between various scenarios. For example, Mastrobuono-Battisti \& Perets (2013) proposed to look for morphological features such as a larger flattening of the enriched population, which is a possible outcome of a scenario with multiple generations of stars (although this may prove challenging to measure). We recently suggested the differential rotation of subpopulations as a promising kinematic signature (H\'enault-Brunet et al. 2015). Given some net angular momentum initially, models for which a second generation forms from gas that collects in a cooling flow into the core of the cluster predict an initially larger rotational amplitude for the polluted stars compared to the primordial stars (a consequence of angular momentum conservation). Hypothetically removing $>90\%$ of first-generation stars (in order to alleviate the ``mass budget" problem plaguing scenarios with multiple generations of star formation) would also remove a large fraction of the angular momentum of the more extended primordial population (see H\'enault-Brunet et al. 2015). The opposite trend is expected from a scenario in which the polluted stars are the ones initially crossing the core of the cluster (e.g. Bastian et al. 2013) and preferentially on radial (low-angular momentum) orbits. In this case, the rotational amplitude of the polluted stars would be lower initially (assuming the cluster has some net rotation). This is illustrated in Fig.~\ref{fig2}, where we also show that a differential rotation between the polluted and primordial stars of the order of 1~km~s$^{-1}$ or larger can remain in the outer parts of a cluster after a Hubble time of dynamical evolution (the simulations shown are scaled to match the properties of 47 Tuc). Marginal evidence for differential rotation between subpopulations has been reported in a few clusters (e.g. NGC 6441 and NGC 6388; Bellazzini et al. 2012), with polluted stars displaying a smaller rotational amplitude, but these are based on relatively small datasets and will need to be confirmed with more observations.

\section{Mass segregation and the spatial distribution of subpopulations in the centre of GCs}

Larsen et al. (2015) found a surprising result when investigating the spatial distributions of subpopulations in M15 using HST/WFC3 photometry of red giants. Unlike what is typically observed in the outer regions of clusters (including M15), they found that the giants with primordial composition are more centrally concentrated in the inner parts of the cluster (within the half-light radius). They considered mass segregation as a possible culprit for the observed trend, but showed with $N$-body simulations that a significant mean mass difference would be needed, with primordial giants required to be $\sim0.25$~M$_{\odot}$ more massive than their enriched counterparts. If interpreted as a difference in helium content, this mean mass difference would imply extreme He enhancement ($Y \gtrsim 0.40$), which is clearly incompatible with the colour-magnitude diagram of M15. Larsen et al. therefore concluded that the radial trends seen in M15 presumably reflect initial conditions, but we consider below another possibility. 

\begin{figure}[h]
\begin{center}
 \includegraphics[width=5.2in]{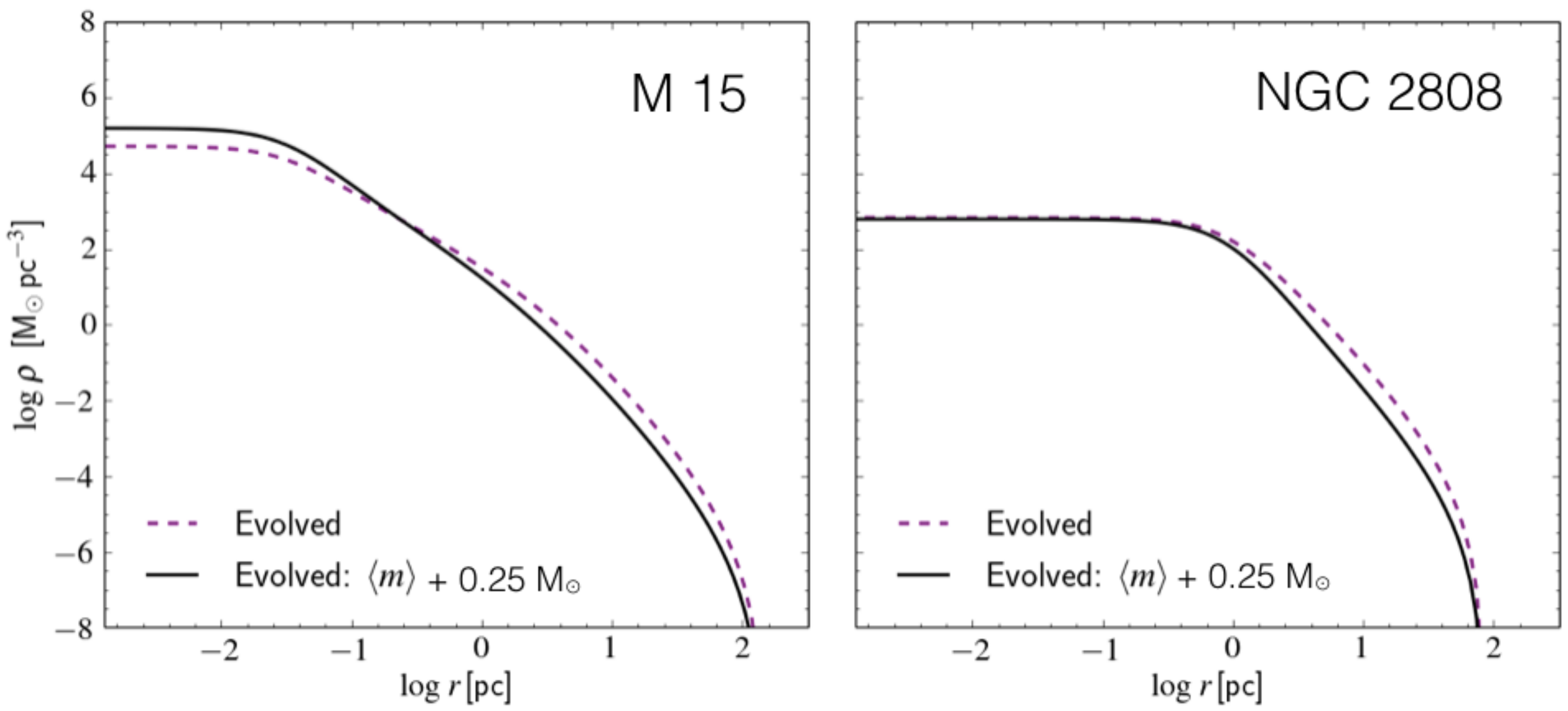} 
 \caption{Predicted mass density profile of two tracer populations of evolved stars with a mean stellar mass differing by 0.25~M$_{\odot}$ in GCs M15 and NGC 2808.}
   \label{fig3}
\end{center}
\end{figure}

A larger binary fraction has been inferred for primordial stars compared to enriched stars ($\sim5-12\%$ vs. $\sim1\%$; D'Orazi et al. 2010, Lucatello et al. 2015) around the half-mass radii of GCs, in keeping with the idea that the enriched stars were born in a denser environment where (loose) binaries are more efficiently destroyed (e.g. Hong et al. 2015). In the core of clusters, the binary fraction can increase to be as high as $40-50\%$ (e.g. Hurley et al. 2007), so this may account for the large mean mass difference needed to explain the results of Larsen et al. (2015). Whether or not this can work in detail however remains to be shown, and it will depend on the complex interplay between many factors: the initial binary properties and spatial distribution of subpopulations, the rate of creation and disruption of binaries, the amount of dynamical evolution, the structure of the cluster. Studies of the spatial distribution of subpopulations in the central regions of other GCs should also help to clarify this picture, but we should keep in mind that the perceived effect of mass segregation on the distribution of different mass species will vary from cluster to cluster. For example, this effect would be stronger in more concentrated clusters like M15 in comparison to clusters with a very large core over half-light radius like NGC 2808. We illustrate this in Fig.~\ref{fig3} by showing, for M15 and NGC 2808, the mass density profile (in 3D) predicted for two tracer populations of evolved stars with a mean stellar mass differing by 0.25~M$_{\odot}$, when assuming a fully mass segregated cluster. The underlying potential was constrained by fitting multi-mass models from Gieles \& Zocchi (2015) to the surface brightness and velocity dispersion profiles of these clusters in order to model their mass distribution.

\section{Summary}

We emphasised that observations of the kinematics and spatial distribution of multiple populations around the half-mass radius of GCs and beyond are consistent with the enriched population forming more centrally concentrated. In contrast, the trend seen in the inner regions of M15 appears to clash with this interpretation and remains difficult to explain by mass segregation. We speculated that a larger binary fraction for the primordial population in the centre of M15 may reconcile these findings. We finally suggest possible future avenues to make progress in understanding the formation of multiple populations in GCs through their long-term kinematics and dynamics: (1) constrain any differential rotation of subpopulations, (2) quantify their central spatial distributions in a larger sample of GCs to compare with the puzzling results in M15, and (3) measure the respective binary fractions of subpopulation in the central regions of clusters.

\begin{acknowledgements}
\noindent{I would like to thank Mark Gieles and Alice Zocchi for useful comments during the preparation of this contribution. I also thank the IAU for a travel grant.}
\end{acknowledgements}

\end{document}